\documentclass[11pt,a4paper]{article}
\usepackage[left=2cm, right=2cm, top=1.50cm, bottom=1.50cm]{geometry}
\usepackage[utf8]{inputenc}
\usepackage{graphicx}
\usepackage{epsfig}
\usepackage{psfrag}
\usepackage{color}
\usepackage{amsmath}
\usepackage{multirow}
\usepackage{amssymb}
\usepackage{latexsym}
\usepackage{amsmath}
\usepackage{balance}
\usepackage{times}
\usepackage{theorem}
\usepackage{pifont}
\usepackage{tabularx}
\usepackage{textcomp}
\usepackage{colortbl}
\usepackage{cite}
\usepackage{icomma}
\usepackage{bbold}
\usepackage{dsfont}
\usepackage{amssymb}
\usepackage{mathptmx}

\usepackage{subfig}
\usepackage{epstopdf}

\newcommand{\Hi}{$\mathcal{H}_{\infty}$}

\newcommand{\K}{\mathcal{K}}

\def\real{\mathds{R}}

\makeatletter
\renewcommand*\env@matrix[1][*\c@MaxMatrixCols c]{%
	\hskip -\arraycolsep
	\let\@ifnextchar\new@ifnextchar
	\array{#1}}
\makeatother

\theorembodyfont{ \normalfont }

\theorembodyfont{ \normalfont }

\usepackage{hyperref}
\usepackage{breakurl}

\begin{document}
\begin{flushleft}
Proceedings 	of the {\em XIV IEEE Latin American Summer School on Computational Intelligence~(EVIC): 12-13-14 December
	USACH Santiago-Chile, 2018}.  \href{http://evic2018.usach.cl/#/}{[evic2018.usach.cl]}.
\end{flushleft}
\vspace{6 mm}

\begin{center}
\textbf{	{\LARGE   Protocol for Energy-Efficiency using Robust Control\\ on WSN\footnote{Supported by the Brazilian agencies CAPES, CNPq,  FAPESP (Grant number 2014/22881-1) and \textit{Departamento de Ingenier{\'i}a Inform{\'a}tica} (DINF) the Catholic University of the Most Holy Conception.}} } 
\end{center}
\vspace{3 mm}		
\textbf{Francisco J. Uribe,$^\star$  Cec{\'i}lia F. Morais,$^\ast$ Jonathan M. Palma$^\ast$}\\

$^\star$ BS.~Students, Departamento de Ingenier{\'i}a Inform{\'a}tica, Universidad Cat{\'o}lica De La Sant{\'i}sima Concepci{\'o}n--UCSC, Chile. 
	{\tt\small fjuribe@ing.ucsc.cl}.
	
$^\ast$ Postdoc researcher and 
	PhD. Student, School of Electrical and Computer 
	Engineering, University of Campinas -- UNICAMP, Brazil. 
	{\tt\small 
		\{cfmorais, jpalmao\}@dt.fee.unicamp.br}.\\

{ \color{blue}
\textbf{ \em Abstract} \textbf{-- The present work analyzes the feasibility of obtaining a single controller (robust), with theoretical guarantees of stability and performance, valid for a total set of network configurations in   designed  the controller for an uncertain success probability obtain the protocol for Energy-Efficiency in Networked Control System NCS. 
In particular, this work investigates the   performance degradation,  in terms of the \Hi~guaranteed cost, between optimal controller design (precisely known probability) and the sub-optimal controller design (robust to probability uncertainties).
The feasibility of the proposed methodology is validated by a numerical example.}\\

\textbf{{\em keywords:  uncertain MJLS,  Energy-Efficiency, NCS}}
}	
			
\section{Introduction}

A full-reliable communication, independent of the network topology, corresponds to the most expensive configuration at the resource level.
However, a full-reliable communication is necessary to implement  the classical control  design  for dynamic systems.
To improve the energy-efficiency in Network Control Systems (NCS)~\cite{HNX:07}, particularly in multi-hop networks used to transmit packet of measurements or control signals, there exists in the literature an approach called {\em Trade-Off} in dynamic systems~\cite{PCGGO:15,PCGGO:16}.
The {\em Trade-Off} approach, for a particular network configuration, provides resource savings in wireless networks,  admitting the employment of semi-reliable networks, because it allows a certain level of performance degradation associated to the controller{\color{blue}~\cite{PCG:15}}.

The protocol for energy-efficiency proposed in~\cite{PDCMORH:18} extends the {\em Trade-Off} approach to deal with time-varying network configurations. Such protocol modifies the maximum number of transmissions
based on of the modification to the maximum number of transmissions per packet (MNTP) per node,
in function of the current energy level of each wireless unit (node).
To confine the MNTP in a bounded interval generates an impact on the probability of successful transmission per package
between the plant and  controller.
A probability of successful transmission  between source and sink depends on the current power   of the wireless units (battery level in the node)~\cite{PDCMORH:18}, the number of possible configurations in the  network depends on;
{\em i)} number of nodes that modify their MNTP. {\em ii)} set of values that the MNTP can assume, for node.
The example shown in~\cite{PDCMORH:18} corresponds to a network with  $16$ nodes
that modify their MNTP and
each node can assume two different values of MNTP,
impllying  $2^{16}$  possible  probability of successful transmission.

The closed-loop  system for  is  composed by different operation modes (depending   on the reception or not of the signals)
which can be conveniently modeled by  called Markov jump linear systems (MJLS).
In the literature~\cite{CFM:05}, there exist design conditions for this class of systems that provide theoretical guarantees of stability and performance.
For precisely known systems subject to packet loss, it is possible to design the optimal controller 
using the MJLS theory.
%
Nevertheless,  such controllers are valid only for unique network configuration (probability), which implies that in the example given in \cite{PDCMORH:18} 
the node responsible for calculating the control signal  must store   $2^{16}$ controllers, one for each network configuration.
In terms of computational cost, 
the storage of $2^{16}$ controllers in Wireless Sensor Networks (WSN) can be inviable.

The present work analyzes the feasibility of obtaining a single controller (robust), with theoretical guarantees of stability and performance, valid for a total set of network configurations (distinct probabilities of successful transmissions).
The designed controller is robust for an uncertain success probability~\cite{MPPO:18b}. 
This project is based on 
on a Markov chain whose state transitions are rulled by a polytopic matrix probability, which includes all the possible network configurations.
%
%
In particular, this work investigates the   performance degradation,  in terms of the \Hi~guaranteed cost, between optimal controller design (precisely known probability) and the sub-optimal controller design (robust to probability uncertainties).
%
The feasibility of the proposed methodology is validated by a numerical example where the
$\mathcal{H}_\infty$ guaranteed  cost obtained for robust approach is numerically  identical
to the one computed with optimal controller (precisely known system).
Therefore, for the study case, the employment of a single robust controller provides the same results than using the optimal solution (that requires a storage of $2^{16}$ controllers) when applied to  applications using the energy-efficiency protocol.

\section{Preliminaries} \label{sec:preliminares}
Consider the notation used in~\cite{MPPO:18b}, and the system 
\begin{equation}
\hspace{-0.1cm} \mathcal{G} \hspace{-0.1cm} = \hspace{-0.1cm}
\left\{
\begin{aligned} 
x(k+1) &= A(\theta_k) x(k) + B(\theta_k) u(k) + 
E(\theta_k)w(k)\\ 
z(k)   &= C_{z}(\theta_k) x(k) + D_{z}(\theta_k) u(k) +  
E_{z}(\theta_k) w(k) \\
%
%
\end{aligned} \right.
\label{eq:SisPrimal-Comp}
\end{equation}
\noindent
where $x(k) \in \real^{n_{x}}$ is the state vector,  $u(k) \in \real^{n_{u}}$ is the control input, $w(k) \in \real^{n_{w}}$ is the noisy input, and $z(k) \in \real^{n_{z}}$ is the controlled output. 
The state-feedback control law is
$u(k)=Kx(k)$. 
%
%
%
The state-space matrices  depend upon a discrete-time homogeneous Markov chain $ \theta_k$ which is associated to a uncertain transition probability matrix given by 
$\Gamma(\alpha) = [p_{ij}(\alpha)]$, $\forall i, j \in \K$,  where $
p_{ij} (\alpha) = \mbox{Pr} \left( \theta_{k + 1} = j ~ | ~ 
\theta_k = i \right), ~~ \forall k \geq 0 $, satisfying $p_{ij} (\alpha) \geq 0$ and $\sum_{j=1}^{\sigma}p_{ij}(\alpha)=1$.
In the study case, 
the  random variable $\theta_k$ has a generalized Bernoulli distribution~\cite{SS:05}, which means that $\Gamma(\alpha)$ has $\sigma$ identical rows or  $p_{ij}(\alpha)=p_j(\alpha)$, $\forall i,j \in \mathds{K}$, more details   in~\cite{MPPO:18b}.
As performance criterion, this work uses the \Hi~norm ($\| \mathcal{G} \|_\infty$). One of the possible definitions for this norm is a  ratio between the expected values of the exogenous input $w(k)$ and the output $z(k)$ for the worst case scenario of the signal $w(k) \in \mathcal{L}_2$~in \cite{CdV:96}, and reproduced below,
\begin{equation} {\footnotesize
	\| \mathcal{G} \|^2_\infty = \sup_{0 \neq w \in \mathcal{L}^2,\ \theta_0 \in \mathbf{K}}  \frac{\|z(k)\|^2_2}{\|w(k)\|^2_2}.}\label{equ-markovian-infinity-norm}
\end{equation}

\section{Main Results: robust {\em vs} optimal $\mathcal{H}_\infty$ Controller}

\begin{figure}[!htb]
	\begin{center}
		\psfragscanon
		\psfrag{q1}[c][c][1][0]{$Q_{in}^1$}
		\psfrag{q2}[c][c][1][0]{$Q_{in}^2$}				
		\psfrag{cc}[c][c][1][0]{$C_c(t)$}				
		\psfrag{cd}[c][b][1][0]{$C_d$}				
		\psfrag{H1}[c][c][1.5][0]{$H_1$~}
		\psfrag{H2}[c][c][1.5][0]{$H_2$}		\subfloat[\label{fig:ex02a}]{\includegraphics[width=0.5\columnwidth]{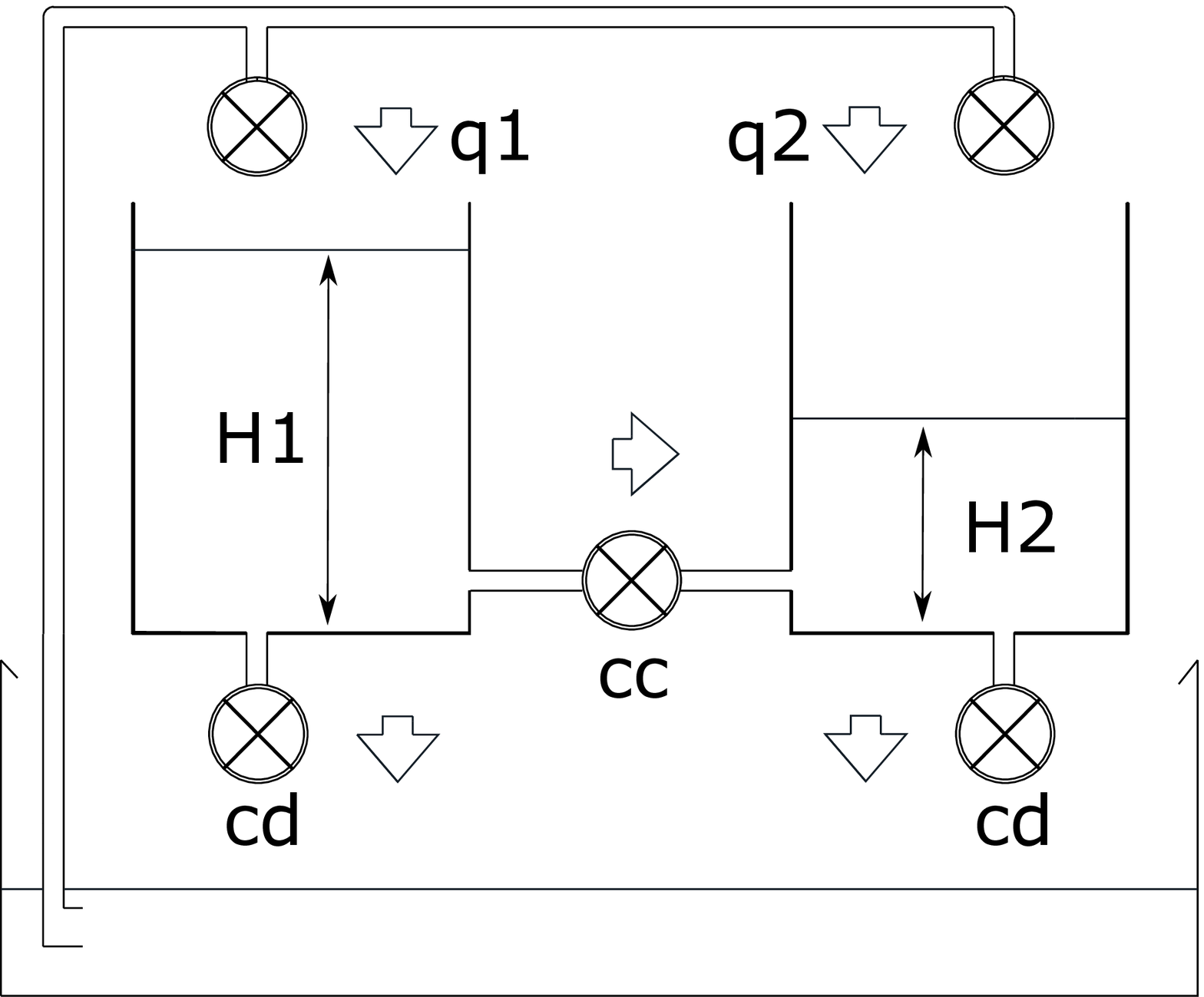}}
		~		
		\subfloat[\label{fig:ex02b}]{\includegraphics[width=0.39\columnwidth]{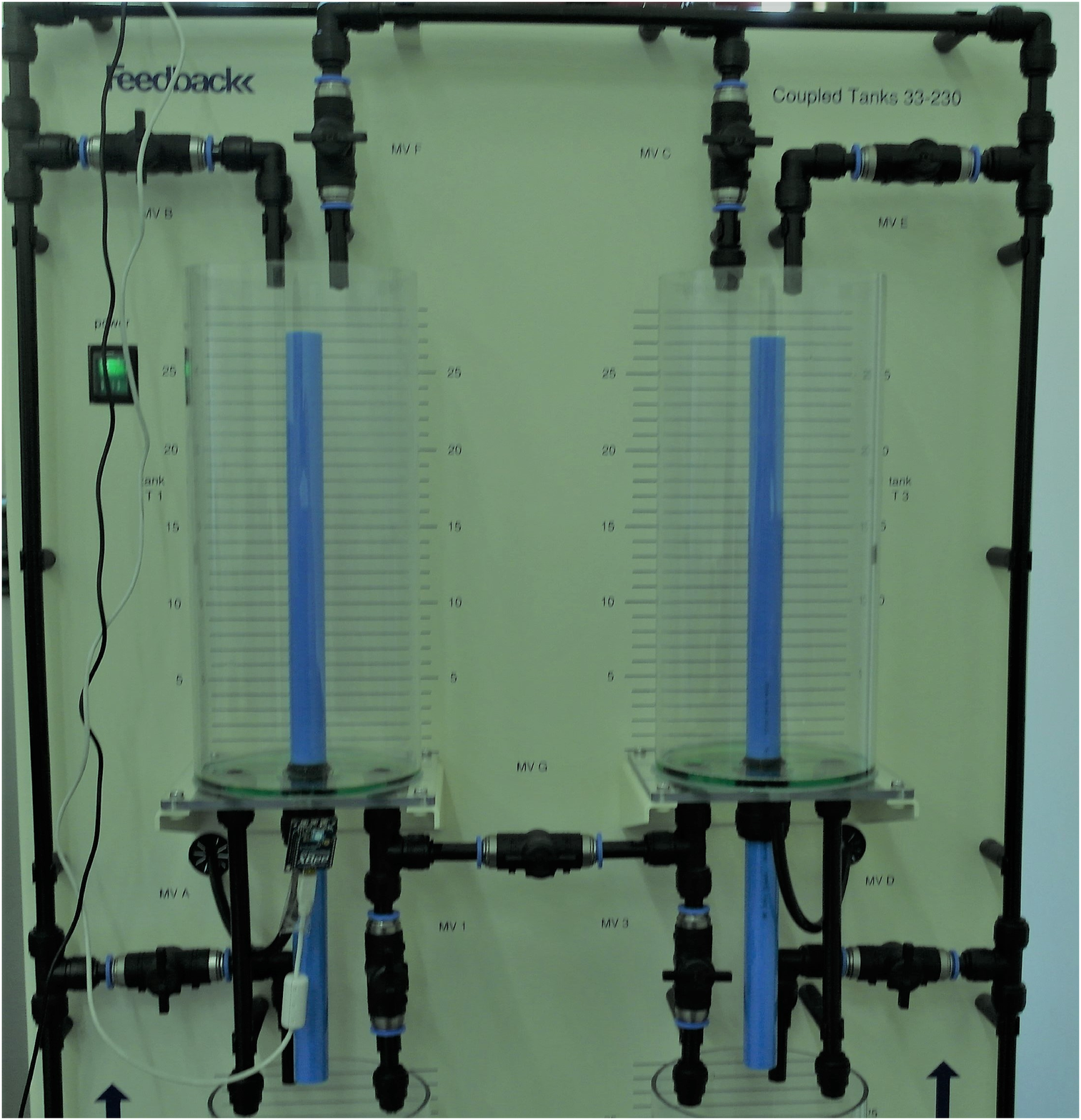}}
		
		\caption{\label{fig:ex02}
			(a) Operating system diagram of Example; (b) Level control plant used in Example.
		}
	\end{center}
\end{figure}

The performance is quantified by the \Hi~guaranteed cost associated with the  optimal controller ($\mathcal{H}_\infty^{OP}$) and with  robust controller ($\mathcal{H}_\infty^{ro}$).  
The \Hi~guaranteed cost for  optimal control design corresponds to actual system norm.
On the other hand, the robust control design provides an upper bound for the \Hi~norm of the system (\Hi~guaranteed cost), 
being of interest the distance 
of the robust with respect to the optimal.
To ilustrate this,
it is used an example of 
level control plant composed of two coupled tanks as shown  in Fig.~\ref{fig:ex02} (borrowed from \cite{PDCMORH:18}).    
The controlled output $(z(k))$ is a level variation of tanks $1$ and $2$, the  physical parameters and space-state matrix for the system used in this example are given in \cite[Seccion 6.1]{PDCMORH:18}.
%


The state-space representation \eqref{eq:SisPrimal-Comp} of the system given in Fig.~\ref{fig:ex02a} is obtained
according to the steps shown in~\cite[Example 2]{PMO:18c} and \cite[Section 4.2]{PDCMORH:18}. 
The system~\eqref{eq:SisPrimal-Comp} models the loss of the control signal $u(k)$ 
by the approach Zero-Input~\cite{Sch:09}, forming two operation modes:  a 
nominal system and a system with loss in the control signal represented by a matrix $B(\theta_k)$ with null elements.
\subsection*{$\mathcal{H}_\infty$ guaranteed cost}
\begin{figure}[htb]
	\begin{center}
		\psfragscanon
		\psfrag{hif}[c][c][2][0]{$\mathcal{H}_\infty$}				 		
		\psfrag{qq}[c][c][2][0]{$q$}					 		
		\psfrag{data1~~}[c][c][1.5][0]{$\mathcal{H}_\infty^{ro}(q)$}
		\psfrag{data2~~}[c][c][1.5][0]{$\mathcal{H}_\infty^{OP}(q)$}
		\includegraphics[width=1\columnwidth]{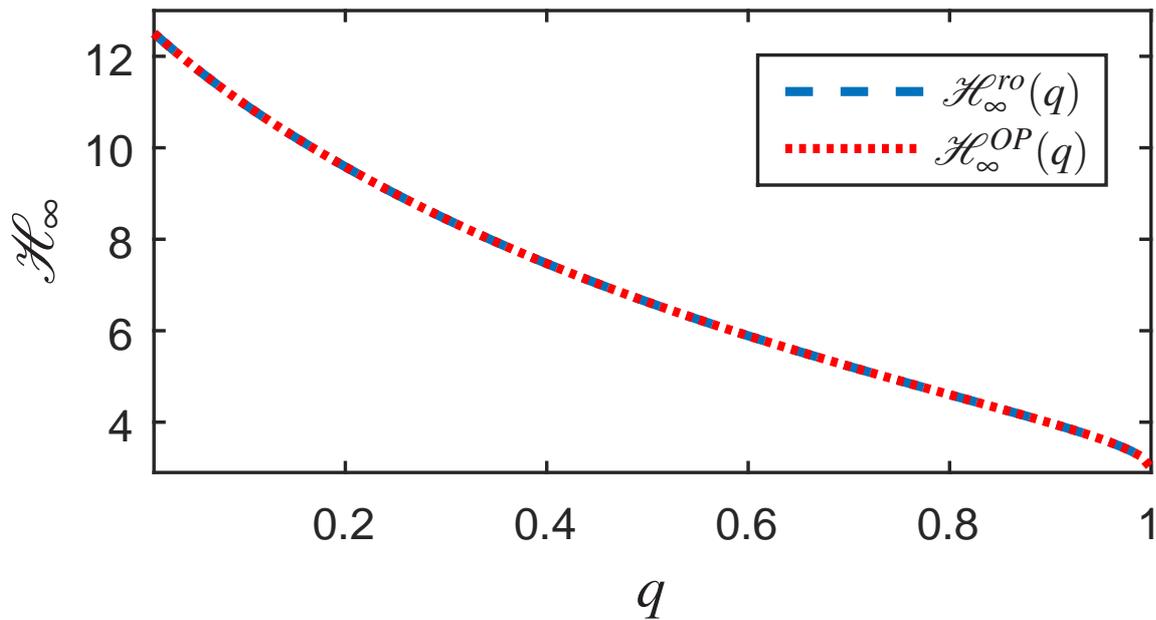}
		\caption{$\mathcal{H}_\infty$ guaranteed cost for the robust and optimal controller in function os $q$.}
		\label{fig:Norm}
	\end{center}
\end{figure}

The Fig.~\ref{fig:Norm} shows the values of:
the \Hi~guaranteed cost ($\mathcal{H}_\infty^{OP}(q)$) of the closed-loop system with the optimal control (precisely known probability) obtained from \cite[Theorem 1, Corollary 1 and $\zeta=10^4$]{MPPO:18b}, in function of the probability q of successful transmission of the control signal $u(k)$.
Fig.~\ref{fig:Norm}   also displays the  \Hi~guaranteed cost ($\mathcal{H}_\infty^{ro}(q)$) of the closed-loop system with the robust controller (polytopic probability matrix $P_S^r(\alpha)$ obtained from \cite[Theorem 1 and $\zeta=10^4$]{MPPO:18b} in function of $q$.
In this case the probability of successful transmission of the control signal $u(k)$ is $P_S^r(\alpha) \in [1, \ q]$.
%
%
%
%
%
%
%
Note that  both the optimal  and the robust cases are numerically equal, implying that, in terms of the \Hi~guaranteed cost,  the design of a robust controller does not add conservatism to the studied problem.
\section{Conclusion}\label{sec:Conc}
The example proposed in \cite{PDCMORH:18} that employs Energy-Efficiency protocol based on WSN  yield $2^{16}$ possible network configurations, requiring the same number of optimal controllers to obtain a certain \Hi~performance level.
In the present work, for the study case, it is shown that a single robust controller (sub-optimal design) can provide the same \Hi~guaranteed costs without incorporate any conservatism.
For resource-constrained networks, as WSN, the robust design presents two advantages: it eliminates the necessity of storing a hard number of controllers and 
also avoids the real-time monitoring of the current network configuration required to switching the controller.
In this case, even a worse performance is acceptable, but in the case study, the improvements were obtained without loss of performance.
%

{\color{blue}
\section{Acknowledgments}
Jonathan M. Palma,  supervisor of the work, acknowledges the \textit{Departamento de Ingenier{\'i}a Inform{\'a}tica} of the Catholic University of the Most Holy Conception
for providing the technical support for the development of the work during his visit between May
  and October 2018. Funded by  PhD Hugo O. Garces, project DINREG 10/2017.}


%
%

\end{document}